\documentclass[fleqn,usenatbib]{mnras}

\usepackage{newtxtext,newtxmath}

\usepackage[T1]{fontenc}
\usepackage{ae,aecompl}

\usepackage{graphicx}	
\usepackage{amsmath}	
\usepackage{amssymb}	

\title[Superbubbles in the Antennae]
{Physical properties of superbubbles in the Antennae galaxies}

\author[A. Camps-Fari\~{n}a et al.]{A. Camps-Fari\~{n}a$^{1,2}$\thanks{E-mail:artemic@iac.es}, 
J. Zaragoza-Cardiel$^{3}$, J.E. Beckman$^{1,2,4}$, J. Font$^{1,2}$,\newauthor P. F. Vel\'{a}zquez$^{5}$, A. Rodr\'{i}guez-Gonz\'{a}lez$^{5}$, M. Rosado$^{3}$\\
$^{1}$Instituto de Astrof\'{i}sica de Canarias, C/V\'{i}a L\'{a}ctea s/n, E-38205 La Laguna, Tenerife, Spain\\
$^{2}$Department of Astrophysics, University of La Laguna, E-38205 La Laguna, Tenerife, Spain\\
$^{3}$Instituto de Astronom\'{i}a, Universidad Nacional Aut\'{o}noma de M\'{e}xico, Apartado Postal 70-264, CP 04510 M\'{e}xico, D. F., M\'{e}xico\\
$^{4}$CSIC, 2806 Madrid, Spain\\
$^{5}$Instituto de Ciencias Nucleares, Universidad Nacional de M\'{e}xico, Apartado Postal 70-264, CP 04510 M\'{e}xico, D. F., M\'{e}xico
}

\begin{document}

\pagerange{\pageref{firstpage}--\pageref{lastpage}} \pubyear{2002}
\maketitle
\label{firstpage}

\begin{abstract}
Mass outflow generated by the dynamical feedback from massive stars is currently a topic of high interest. Using a purpose-developed analysis technique, and taking full advantage of the high kinematic and angular resolution of our instrument we have detected a number of expanding superbubbles in the interacting pair of galaxies Arp 244 (NGC 4038/9) commonly known as the Antennae. We use a Fabry-P\'{e}rot interferometer GH$\alpha$FaS to measure the profile of H$\alpha$ in emission over the full extent of the object, except for the extended HI tails. The superbubbles are found centred on most of the brightest HII regions, especially in the overlap area of the two merging galaxies. We use measured sizes, expansion velocities and luminosities of the shells to estimate most of the physical parameters of the bubbles, including the kinetic energy of the expansion. In order to assess the validity of our results and approximations we perform a hydrodynamic simulation and manage to reproduce well our best measured superbubble with reasonable physical input assumptions. We also study the sources of ionization of the shells, finding that at the current, quite late stage of expansion, radiation from the remaining stars dominates, though the effect of supernova shocks can still be noted.

\end{abstract}


\section{Introduction}

Massive stars have a profound impact on the surrounding interstellar medium (ISM), their winds and supernovae carving out the surrounding volume, creating shells, cavities, and outflows of mass from the galaxy disc. This effect is amplified by the presence of these stars in OB associations. In those cases a significant quantity of gas can be blown out into the galactic halo, and in stronger outflows can escape into the intergalactic medium. The feedback from massive stars is critical for the self-regulation of star formation, which can be quenched when the parental molecular cloud is disrupted, or even enhanced if the expanding gas destabilizes nearby molecular clouds, which collapse to produce new stars \citep{Gerola1978,McCray1987,Palous1994}. This feedback is put suggested as the cause of the overall low star formation efficiency observed in spiral galaxies, which extends the duration of star formation and prevents the gas from being depleted too quickly \citep{Dalla2008,Ceverino2009}. A correct modelling of these effects is, according to this scenario, a key to produce realistic simulations of galaxy formation and evolution.

The injection of energy by massive stars into their surrounding medium generally produces expanding bubbles which occur over a wide range of scales, from sizes of a few pc around single stars to kpc scale superbubbles blown by starbursts. Measuring the properties of these bubbles, notably their kinetic energies of expansion, is essential for the evaluation of stellar feedback, in galaxy discs and in their circumnuclear zones. These measurements can also be useful for indirect estimation of the properties of the originating star or star cluster, constraining parameters such as the minimum number of OB stars in a cluster which could power the bubble. It is particularly interesting to estimate the age of the bubble, because this allows us to estimate the age of the star cluster itself, given that feedback commonly shuts off further star formation inside an HII region after the initial burst and the winds by the massive stars begin to act immediately on the gas. However the accuracy of these estimates is limited by the reliability of the model used to predict the evolution of the expanding bubble, which tend to be simplified, unless one performs detailed studies of individual cases, and preferably with well-structured simulations. In spite of the limitations this is a useful method for estimating cluster ages when lacking the resolution needed to resolve the individual stars.

The strongly interacting pair of galaxies NGC 4038/4039 (also Arp 244, and commonly known as the Antennae) is in the intermediate stages of a merger. The interaction has triggered a powerful starburst, notably in the overlap region where the interaction is strongest. As this object is one of the nearest examples of this type of starbursts it attracts considerable attention, and is one of the most studied galaxy pairs in the sky. This environment is an ideal laboratory for the probing of feedback by massive stars.
The young stellar clusters have been studied extensively both as a population, measuring their luminosity and mass functions \citep{Whitmore1999,Fritze1999,Zhang1999,Mengel2005,Whitmore2010}, and individually in detail \citep{Bastian2006,Mengel2002,Whitmore2005}.
There are also studies of the molecular gas and dust, and their relation to the star clusters \citep{Wilson2000, Nikola1998,Gao2001,Zhu2003,Wilson2003}. In a recent publication \citep{Zaragoza2014} the authors of the present paper have compared the mass functions and the density distributions of the molecular and ionized gas clouds, showing that for both gas phases there are two populations, divided at a critical mass which is comparable for the HII region population and that of the molecular clouds.

Bubbles and superbubbles in general have long been subjects of study, as they have been well detected in the Galaxy and in nearby galaxies. Most of these detections have been made in neutral hydrogen, using the kinematics of the 21 cm line, and include the local bubble around the solar neighbourhood \citep{Cox1987} as well as bubbles in M31 \citep{Brinks1986}, M33 \citep{Deul1990}. Detections have also been made using the ionized gas with H$\alpha$ in the Magellanic clouds \citep{Meaburn1980,Meaburn1984,Rosado1986,Le1993,Chu1994}, and more recently \cite{Ambrocio2016,Reyes2014} for the LMC superbubbles. Examples of expansion in galactic outbursts such as in M82 \citep{Bland1988}, NGC 3079 \citep{Veilleux1994} and the irregular galaxy IC 1613 \citep{Valdez2001} have been reported. These are only a few selected examples of a progressively extending literature on the subject. But given the importance of stellar feedback on galaxy evolution in general amplifying the observation and above all the quantification of superbubbles to as large a sample of galaxies as possible is of strong interest to the field. This interest has grown in recent years with several studies pointing to the feedback from massive stars as an important factor in a wide variety of astrophysical problems. These include the possible dissipation of nuclear dark matter cusps \citep{Pontzen2012} the dissemination of metals both within galaxies \citep{Spitoni2009} and into the intergalactic medium \citep{Heckman1990} and the enhancement of the infall rate of low metallicity intergalactic gas to galaxy discs by interaction with supernova ejecta in the galactic halo \citep{Marasco2012}.

\section{Observations}

We observed the Antennae pair of galaxies using the Galaxy H$\alpha$ Fabry-Perot system (GH$\alpha$FaS,\cite{Hernandez2008,Fathi2008}), obtaining a calibrated data cube which mapped kinematically the H$\alpha$ emission line over the extent of both galaxies, with only the extended HI tails falling outside the field. In practice we mapped the full extent of the H$\alpha$ emission from the object. The observations were first presented in \cite{Zaragoza2014}.

GH$\alpha$FaS is an integral field spectrometer mounted at the Nasmyth focus of the 4.2m William Herschel Telescope (WHT) at the Observatorio del Roque de los Muchachos (ORM) La Palma. It produces a 3.4 arcmin x 3.4 arcmin data cube with seeing-limited angular resolution (spaxel size $\sim0.2$ arcsec) and a spectrum over each spaxel with 48 channels covering a spectral range which depends to second order on the wavelength range of the observed emission, but is close to 400 km/s yielding a sampling velocity resolution of $\sim8$ km/s at H$\alpha$.

The data reduction procedures for GH$\alpha$FaS have been well described in the literature (see e.g. \cite{Blasco2010}), so here were give a very brief summary. In order to optimize the field size and the optical throughput GH$\alpha$FaS does not use a field derotator so the required rotation correction is applied a posteriori with software, which is described, along with the procedures for velocity calibration, phase correction, sky background removal and adaptive binning, all of which are described in \cite{Hernandez2008} or \cite{Blasco2010}. The end product of this initial reduction is a data cube which is used here as described below.

We used a flux-calibrated and continuum-subtracted image of the object in H$\alpha$ for the calibration procedure of the shells detected with the data cube, as we have learned from experience that this is more reliable than calibrating the cube directly. The image was taken with the direct imaging camera ACAM, also on the WHT.

For the spectroscopic part of the study we used the Intermediate Dispersion Spectrograph (IDS) at the Cassegrain focus of the 2.5 m Isaac Newton Telescope (INT) also at the ORM. This yields long-slit spectra, 3.3 arcmin unvignetted, and we used a slit-width of 1 arcsec, and a grating which produced a plate scale of 0.44 A/pixel using the RED+2 detector. The aim of the long-slit spectroscopy was to measure the relative intensities of the Ha, [NII] doublet, and [SII] doublet lines, to compute line ratios and then perform tests on the ionizing radiation of the detected expanding shells. The spectra were reduced using IRAF and was particularly simple; the lines are close enough together and observed with the same system so that reddening correction was not needed, nor was calibration necessary as we were interested only in the line ratios.

\section{Expansion map}
\begin{figure}
\includegraphics[width=\linewidth]{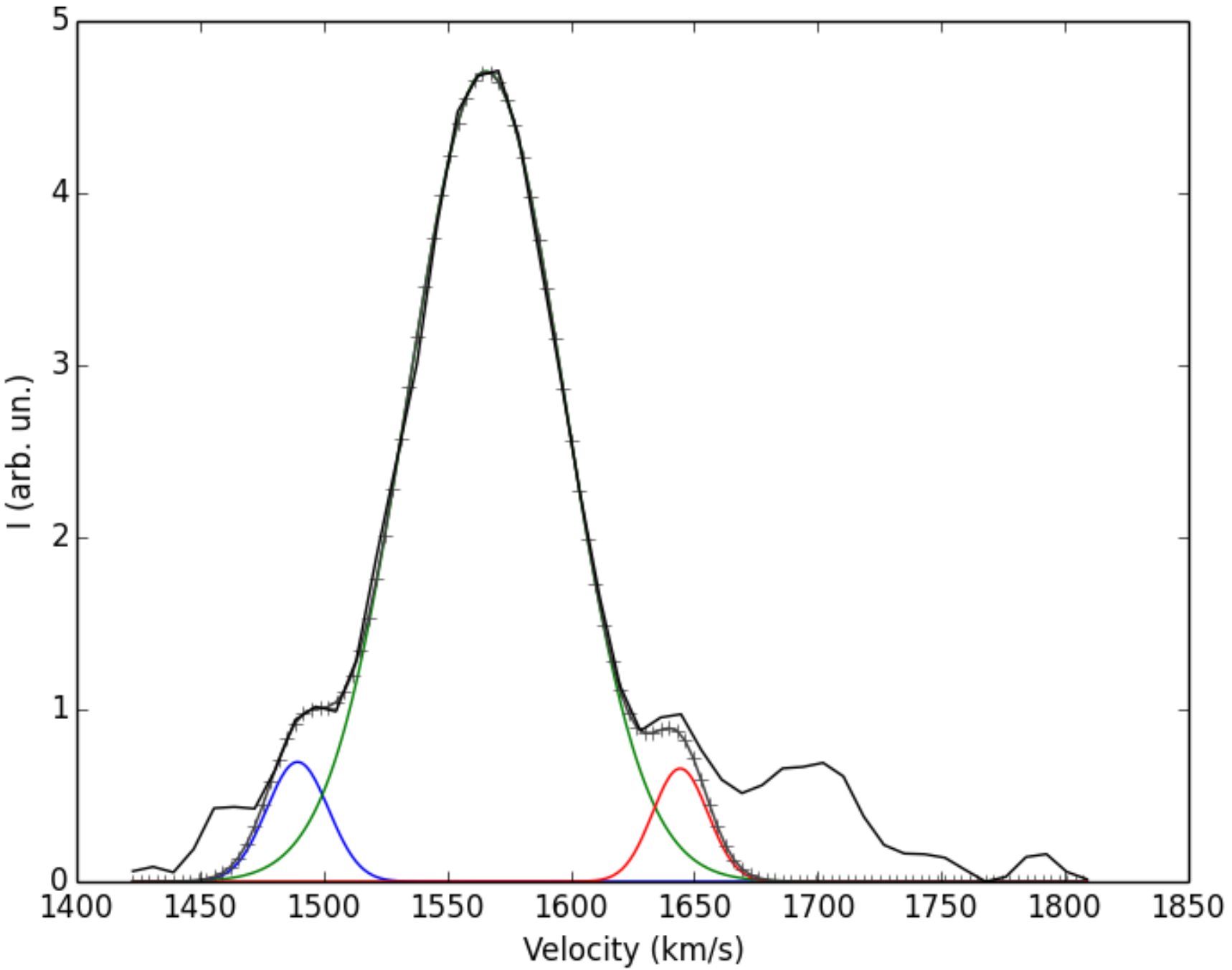}
\caption{Line profile of a spaxel in the data cube showing a clear expansion signature. We plot the data in a black line, with the individual components in color, green for the main peak corresponding to the bulk of the region and blue and red for the approaching and receding components of the shell respectively. It is apparent that the two peaks lie symmetrically spaced from the bulk of the region, the signature of the presence of an expanding shell associated with an HII region. Further peaks can be observed on the profile, but since they do not show a similar correlation in velocity to the former ones we do not consider them.}
\label{fig:exp_ex}
\end{figure}

To detect the superbubbles we used our purpose-developed program \textsc{bubbly}. This is described in detail in \cite{Camps2015} so here we give only an outline of how it works. HII regions are optically thin in H$\alpha$, the only extinction is provided by any dust present. This enables us to observe simultaneously any material expanding towards the observer and any material expanding away from the observer, from the front and the back of an expanding shell respectively. As we observe an ionized zone in H$\alpha$ the emission line profile at any point where there is a bubble along the line of sight is triple-peaked with the brightest, central peak emanating from the bulk of the HII region, and two symmetrically placed fainter peaks emitted by the shell. Our program, \textsc{bubbly}, first detects the presence of multiple components in the line profiles across the full field of the data cube, and then searches for the tell-tale signature of symmetrically blue- and red-shifted (within uncertainty limits) weaker emissions in the wings of a main emission peak. In Fig. \ref{fig:exp_ex} we show an example of a spectrum showing this type of expansion signature (the other weak features may be real emission but not due to a bubble as they are not symmetrically placed with respect to the main peak).

\begin{figure*}
\includegraphics[width=\linewidth]{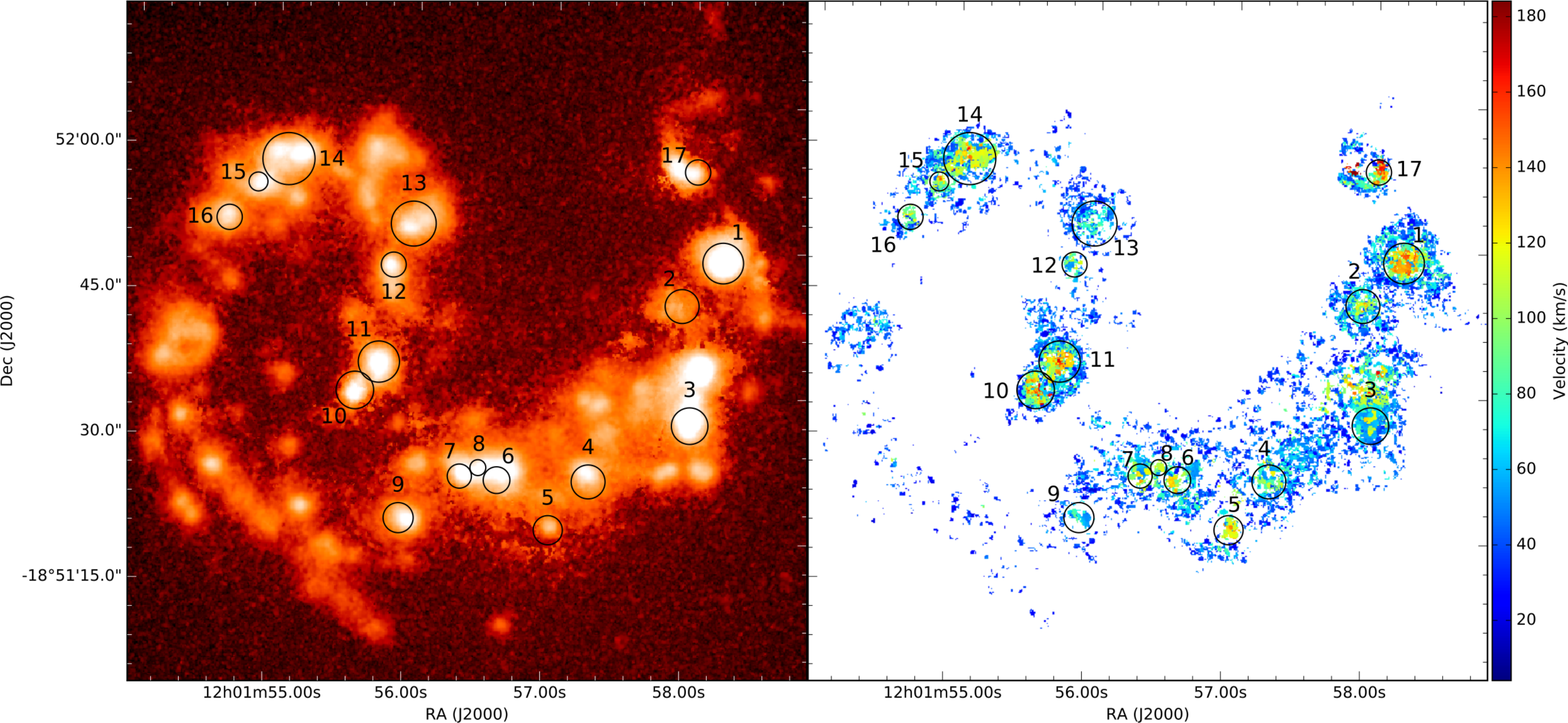}
\caption{On the right we show the expansion map for the Antennae galaxies, colour coded in velocity of expansion, that is, the mean separation in velocity of the secondary components. On the left is a map of H$\alpha$ intensity for comparison. The circles represent the superbubble detections in both maps for ease of comparison, with their radius coinciding with the estimated bubble radius. The superbubbles clearly appear around most of the brightest regions. The presence of an extended,widely disseminated low expansion velocity component can also be appreciated (see text for a discussion on its possible origins).}
\label{exp_map}
\end{figure*}

We represent the information obtained with \textsc{bubbly} in the form of an "expansion map" produced by assigning to each spatial point the mean velocity separation of the secondary peaks detected, if any. We can then use this expansion map to locate bubbles, searching for spatially coherent structures within the detected expansion. This allows us to derive the radial size of the bubble, one its most relevant physical parameters, as well as to discard apparent, but false detections which can arise from noise or from independent clouds coincidentally aligned along particular lines of sight.

In Figure \ref{exp_map} we show the expansion map of the Antennae, next to an H$\alpha$ surface brightness image, for comparison.
The map shows detected expansion at some level of significance across practically the full extent of the area with detected H$\alpha$ emission, but it is clear that around most of the brightest HII regions we find more coherent structures, which show higher detected velocities of expansion. These are what we claim are superbubble detections, and this claim is supported by the coincidence of their centres with those of the HII regions themselves, as well as by the radial symmetry they show. In the figure we have marked the presence of each superbubble with black circles both on the expansion map and on the H$\alpha$ map; these circles show their estimated radial sizes.

Some of these bubbles are very well defined: bubbles with the identifiers 1, 10, 11 and 14 are particularly prominent, showing better coverage and spatial coherence. It is also interesting to pay particular attention to the overlap zone, the area where the interaction has the most effect on the ISM, and where there are some of the brightest HII regions, which have made this the most explored part of the object as found in the literature. In this area we find the clearest superbubble, identified as bubble 1, but we also see in the area below bubble 2 a zone with quite a lot of high velocity expansion which does not, however, show circular morphology coincident with the underlying HII regions. This probably indicates that while there are indeed massive outflows from these young regions they do not maintain the pseudo-spherical shape of a bubble, (due probably to the inhomogeneity of the surrounding ISM), or that the shape is too complicated for us to determine with our data and method. For an irregularly shaped expanding shell the difference in velocity between the approaching and receding surfaces would in any case prevent our method from detecting it. This is a possible factor underlying the incoherent detections shown.

We can also see that, apart from the detections on and around the bright HII regions there is extended, less geometrically coherent detection across the full face of the interacting galaxies. It is interesting to note, however, that in other galaxies analysed in the same way we have found much less, or none of this extended expansion detection, above all in those galaxies in a more quiescent state. This leads us to believe that the dominant cause of these detections is not noise, but unresolved expansion, and more complex kinematic structure. These features also have low expansion velocities, which fits the scenario of smaller, less energetic drivers of the motion.

\section{Physical properties of the superbubbles}
From the expansion map we have a direct measurement of the expansion velocity and size of a superbubble. The flux is calculated using the relative intensities of the secondary peaks and a calibrated H$\alpha$ image of the object. These are the primary quantities used to calculate the other physical parameters. A detailed description of the calculations can be found in \cite{Camps2015}, but essentially we use the relation:

\begin{equation}
L_{H\alpha} (shell) = 4 \pi R^2 \Delta R n_e^2 \alpha (H,T) h \nu
\end{equation}

$R$ stands for the shell radius, $ \Delta R$ is its thickness and $n_e$ is the electron density. $\alpha (H,T)$ is the effective recombination coefficient for hydrogen and $h \nu$ is the energy of an H$\alpha$ photon.
We have no direct way of measuring the shell width $ \Delta R$ so we need to use an approximation. In section 5 we perform hydrodynamical simulations of one of the detected bubbles, from which we derive a canonical value of 15\% of the radius, which is also consistent with observations of local superbubbles \citep{Oey1996}.

With these quantities we can estimate the electronic density in the shell, and together with the radius, the shell thickness, and the velocity we can derive the total mass and the current kinetic energy for the ionized gas. While the uncertainty in the shell thickness certainly affects our results the systematic uncertainty depends on $ \sqrt{\Delta R}$ implying that an error by a factor as high as 3 in shell width implies only a change of $\sim 1.8$ in density, mass and energy. The density decreases with increasing shell width, whereas mass and density decrease. The values of the derived physical parameters for all the identified superbubbles are given in Table \ref{prop_tab}. An additional initial approximation we have used is that the bubbles are spherical. This would be good for homogeneous environments, but as our superbubbles are comparable to the scale height of the surrounding ISM, its density gradient is relevant. Many superbubbles are approaching or in the breakout phase of their evolution, so their shapes will be elongated perpendicular to the disc, i.e. in the direction of falling density, and their expansion velocity will not be isotropic. In section 5 we use hydrodynamic simulations to reproduce one of the superbubbles, which gives us a useful approach to discuss these effects and their implications for the validity of our numerical parameters.

\begin{table*}
\centering
\caption{Table with the physical parameters of the superbubbles detected in the Antennae galaxies.}
\setlength{\tabcolsep}{5pt}
\begin{tabular}{lcccccccll}
No. & Radius & $v_{exp}$ & L$_{shell}$ & $n_e$ & Mass & E$_k$ & Age & Age$_{\mathrm{G\&G}}$ & Age$_{\mathrm{W\&Z}}$\\
 & (pc) & (km/s) &  $(10^{39}$ erg/s) & (cm$^{-3}$) & ($10^{35}$ Kg) & ($10^{51}$ erg) & (Myr) & (Myr) & (Myr)\\
(1) & (2) & (3) & (4) & (5) & (6) & (7) & (8) & (9) & (10)\\

\hline
1 	&	391	$\pm$	39	&	132	$\pm$	13	&	1.45	$\pm$	0.1	&	0.73	$\pm$	0.08	&	31	$\pm$	6	&	267	$\pm$	76	&	2.9	$\pm$	0.4	&	3.45	&	3.8,3.7$^a$	\\
2 	&	326	$\pm$	33	&	115	$\pm$	12	&	0.9	$\pm$	0.1	&	0.77	$\pm$	0.1	&	19	$\pm$	4	&	124	$\pm$	35	&	2.8	$\pm$	0.4	&	5.72$^a$	&		\\
3 	&	352	$\pm$	35	&	62	$\pm$	6	&	1.3	$\pm$	0.1	&	0.83	$\pm$	0.09	&	25	$\pm$	5	&	49	$\pm$	14	&	5.5	$\pm$	0.8	&		&	3.0	\\
4 	&	326	$\pm$	33	&	115	$\pm$	12	&	0.2	$\pm$	0.03	&	0.36	$\pm$	0.04	&	9	$\pm$	2	&	57	$\pm$	16	&	2.8	$\pm$	0.4	&	6.19	&	7.4,2.5	\\
5 	&	280	$\pm$	28	&	125	$\pm$	13	&	0.14	$\pm$	0.02	&	0.38	$\pm$	0.04	&	6	$\pm$	1	&	45	$\pm$	13	&	2.2	$\pm$	0.3	&		&		\\
6 	&	257	$\pm$	26	&	125	$\pm$	13	&	0.22	$\pm$	0.04	&	0.53	$\pm$	0.07	&	6	$\pm$	1	&	49	$\pm$	14	&	2	$\pm$	0.3	&		&	8.4$^a$	\\
7 	&	232	$\pm$	23	&	120	$\pm$	12	&	0.21	$\pm$	0.03	&	0.61	$\pm$	0.08	&	5	$\pm$	1	&	38	$\pm$	11	&	1.9	$\pm$	0.3	&		&	5.0	\\
8 	&	153	$\pm$	15	&	110	$\pm$	11	&	0.09	$\pm$	0.02	&	0.7	$\pm$	0.1	&	1.8	$\pm$	0.4	&	11	$\pm$	3	&	1.4	$\pm$	0.2	&		&	4.0$^a$	\\
9 	&	289	$\pm$	29	&	97	$\pm$	10	&	0.18	$\pm$	0.02	&	0.41	$\pm$	0.05	&	7	$\pm$	1	&	32	$\pm$	9	&	2.9	$\pm$	0.4	&		&	2.0$^a$	\\
10	&	361	$\pm$	36	&	152	$\pm$	15	&	0.92	$\pm$	0.1	&	0.66	$\pm$	0.07	&	22	$\pm$	4	&	250	$\pm$	71	&	2.3	$\pm$	0.3	&	7.0	&	2.0	\\
11	&	396	$\pm$	40	&	168	$\pm$	17	&	0.37	$\pm$	0.03	&	0.36	$\pm$	0.04	&	16	$\pm$	3	&	224	$\pm$	63	&	2.3	$\pm$	0.3	&		&	2.0,4.8$^a$	\\
12	&	240	$\pm$	24	&	115	$\pm$	12	&	0.12	$\pm$	0.02	&	0.43	$\pm$	0.05	&	4.2	$\pm$	0.8	&	28	$\pm$	8	&	2	$\pm$	0.3	&		&	2.0	\\
13	&	432	$\pm$	43	&	85	$\pm$	9	&	0.4	$\pm$	0.04	&	0.33	$\pm$	0.04	&	19	$\pm$	4	&	68	$\pm$	19	&	5	$\pm$	0.7	&		&	6.0	\\
14	&	501	$\pm$	50	&	117	$\pm$	12	&	1.3	$\pm$	0.2	&	0.48	$\pm$	0.06	&	42	$\pm$	8	&	287	$\pm$	81	&	4.2	$\pm$	0.6	&		&	4.5,6.4,4.0$^a$	\\
15	&	182	$\pm$	18	&	145	$\pm$	15	&	0.12	$\pm$	0.02	&	0.67	$\pm$	0.08	&	2.8	$\pm$	0.6	&	30	$\pm$	8	&	1.2	$\pm$	0.2	&	6.61$^a$	&	7.0	\\
16	&	243	$\pm$	24	&	120	$\pm$	12	&	0.17	$\pm$	0.03	&	0.51	$\pm$	0.06	&	5	$\pm$	1	&	37	$\pm$	10	&	2	$\pm$	0.3	&		&	5.2$^a$	\\
17	&	243	$\pm$	24	&	165	$\pm$	17	&	0.24	$\pm$	0.04	&	0.61	$\pm$	0.08	&	6	$\pm$	1	&	83	$\pm$	23	&	1.4	$\pm$	0.2	&		&		\\

\end{tabular}

\textbf{Notes.} Col. (1) corresponds to the superbubble identifier numbers, radius (2) and $v_{exp}$ (3) are estimated directly on the expansion map and have an assigned 10\% error. Luminosity is measured from the ACAM image, its error inferred from the difference in luminosity at low and high limits for radius. The rest of the properties are calculated using the first three (see text for details) except for the cluster ages (9,10) which are taken from spectroscopic and photometric studies of the clusters. Column (9) corresponds to ages measured by \cite{Gilbert2007} using their Br-$\gamma$ equivalent width. Column (10) shows \cite{Whitmore2002}'s measurement of the cluster ages, comparing UBVI colors with spectral evolution models. The clusters are matched spatially to our bubbles, though several bubbles encompassed several clusters, we show the ages for all spatially coincident clusters, ordered by putting the clusters closest to the centre of the bubble first. The less reliable matches, which correspond to clusters found close to the edge of the superbubble detection are marked with $^a$.
\label{prop_tab}
\end{table*}

The age we present is a simple estimate, obtained by dividing the observed radius by the expansion velocity, as though the bubbles expanded freely. The surrounding gas will always slow down the bubbles, so these estimates are lower limits to the bubble ages. A formula which has been commonly used in the literature for the expansion time t, is $t = 0.6R/v$, which is obtained assuming constant momentum injection into a homogeneous isotropic medium. But this is a reasonable assumption only for bubbles which are small compared to the disc scale height, which is not the case for the superbubbles we have found here. As there is no standard treatment of this problem, we have preferred, in the first instance to use fewer assumptions and present lower limits to the ages.

Fortunately, as this is a well-studied object, there are references in the literature for the ages of the star clusters in this object. \cite{Gilbert2007} use Hubble Space Telescope (HST) spectra to measure the Br-$\gamma$ equivalent width of the clusters and compare it to synthetic stellar populations obtained using \textit{Starburst99}. Another source is \cite{Whitmore2002}, where they improve on a previous age determination comparing \textit{UBVI} colors to spectral evolution models by A.G. Bruzual and S. Charlot. This is done using HST H$\alpha$ images to break the age-reddening degeneracy and correct the contamination of the V observations by the emission line.
We present the associated cluster ages in Table \ref{prop_tab} for each bubble. The clusters were spatially matched to the measured extent of the superbubbles, and all matches found are presented for each age determination. Given the size of our superbubbles, they tend to encompass several stellar clusters, so in some cases we have multiple age determinations associated with a superbubble. Some of these lie close to the edge of the superbubble, making them less reliable, these are indicated in the table.
Most of our superbubbles have at least a cluster with similar age, with only bubbles 5 and 17 having no clear spatial match. This does not mean that there is no detected underlying cluster, only that there is not an age determination available from the literature.

The superbubbles listed in Table \ref{prop_tab} are all very large, with radii in the range 150-500 pc. We do not detect smaller bubbles because of an inherent selection effect caused by the limit to the linear resolution at the distance of the Antennae and our requirement of morphological coherence to admit bubble candidates. As explained above, the smaller bubbles should be present within the widespread expansion signatures detected across the whole of the merging discs. Indeed there are "detection clusters" with 3-8 conjoined pixels having similar expansion velocities, but too small for us to make an estimate of a radius with reasonable accuracy. Given that the inferred physical parameters of a bubble are strongly dependent on its radius, we have chosen not to include these as detected bubbles in our list but have tried to make a macroscopic quantification of the total kinetic energy injected into the ISM of the complete system in a following subsection.

The luminosities of the bubbles correlate well with the total luminosity of the region in the area occupied by the bubble; the bubble luminosity is around an order of magnitude less than the luminosity of its region. Bubble 1 is significantly brighter than the others, with correspondingly higher values for the mass and kinetic energy of its shell, as might well be expected as its associated HII region is by far the brightest and youngest of the giant HII regions in the Antennae. The kinetic energy of an expanding shell is its key parameter, given the important role played by the energy injected into the ISM in galaxy evolution. It also provides a conservative constraint on population of the cluster, giving an estimated lower limit to the number of massive stars in the cluster necessary to power the expansion. To make interpretation easier we present the kinetic energies of the shells in units of $10^{51}$ erg, a canonical value of the kinetic energy produced in a supernova explosion. This is not to be taken as an estimate of the number of supernovae that have powered the superbubble, because the energy losses during the expansion are quite high, and increase in time as the shell moves outwards and each supernova remnant has to travel further to reach the shell.

\subsection{Extended detections}
In order to estimate the kinetic energy in the unresolved expansion we carried out an estimate, using a set of assumptions. The main problem in calculating the kinetic energy associated with the unresolved detections is the value for the expansion radii. The velocity and luminosity detected on each pixel are straightforward quantities in the sense that they can be assigned to sections of a bubble without much loss of generality save for the projection effects on the velocity. The radius, however, is a global property of the bubble, so lacking a measurement of the radius corresponding to each pixel we cannot evaluate it correctly. For this reason we will assume that for our purposes each pixel is composed of a slab of material of dimensions X, Y, Z where X and Y correspond to the pixel size of 0.2 arcsec, corresponding to 23.4 pc, and Z is taken as the width of a shell of typical size, assuming a width of 15\% as for the resolved superbubbles. In order to obtain a "typical" bubble size we use \cite{Zaragoza2015}'s sample of HII regions \footnote{Available at Vizier: J/MNRAS/451/1307} and take the median radius of the HII regions with radius lower than that of our smallest resolved bubble (153 pc). In \cite{Zaragoza2015} the HII regions with an uncertainty in radius greater than 15\% are discarded, but we need not do this for the estimate sought here. This selection would bias the sample to higher radii, and as we want the median value the individual uncertainties will cancel. The resulting median radius is 60 pc. Using this value the total kinetic energy for the extended, unresolved detection s is $1.72 \cdot 10^{54}$ erg, while the total equivalent energy in the resolved bubbles is $1.68 \cdot 10^{54}$ erg. The energy in the extended detections is thus of the same order as that in the resolved bubbles.

A possible caveat for this comparison is that the method for calculating the energy is different from that used for the resolved bubbles, where we have been able to measure the derived radius and expansion velocity. To account for this we simulated the effect by calculating the ratio between the two methods for a 60 pc radius bubble at our spatial resolution.

To project the bubble emission into equivalent pixels we calculated geometrically the volume of the intersection between a 23.4 pc x 23.4 pc wide column and a spherical shell to take into account the quantity of gas which emits the light observed in a single pixel. This enables us to derive a mock luminosity at the resolution of the instrument which, combined with the projected velocity along the line of sight allows us to estimate the difference between the true kinetic energy of the simulated bubble and that measured by summing the energy over all the pixels. For a bubble of typical size 60 pc the real energy is 1.9 times higher than the value calculated in this exercise. We carried out the equivalent computation for the full range of sizes in the HII region sample, and the median ratio coincides with that for the median radius, yielding a value of 1.9 This result implies that the method used to estimate the kinetic energy for the extended detections does not entail a large discrepancy from the true value, so if we correct by the factor the energy in the extended detections is $3.27 \cdot 10^{54}$ erg, now somewhat larger than, but still of the same order as that for the resolved superbubbles.

Of course this value cannot be taken as an accurate measurement, as there are several competing effects which could lower or raise it. We would expect the typical value for the radius to be skewed to larger radii because of the selection due to the limit of spatial resolution, which we would mean that we are overestimating the shell width, and consequently the mass and the energy. However our detections of expanding features are almost certainly not complete. The SNR ratio on which we base the estimates affects the fraction of true detections which are discarded, as well as false noise-generated "detections" which are included.

Even with these uncertainties it is apparent that the budget for kinetic energy injection into the ISM in this merging pair of galaxies is not dominated by the giant HII regions; the smaller regions are essentially on a par. This kind of energy input, affecting the whole of the observed pair of merging galaxies, must be taken into account when modelling the physics of mergers.

\section{Hydrodynamic simulation}
As the bubbles we have detected in the Antennae have radii of of order several hundred pc we would expect them not to be spherical but to expand more rapidly in the direction perpendicular to the plane of the disc of their originating galaxy, and therefore to be elongated in that direction \citep{Mac1988} even though the discs have become somewhat distorted during the interaction. They are likely to be approaching the break-out phase. In order to better quantify the values of the parameters derived from our observations, notably the ages of the bubbles, we considered it valuable to run models to simulate the phenomena. The exercise we used to test this approach was to attempt to reproduce our best detection, namely bubble 1.

\subsection{Modelling setup}
To perform the simulation we used a version of YGUAZU-A \citep{Raga2000,Raga2002} a 3D adaptive grid code which solves the Euler equations for each grid element. It was created in order to simulate the conditions of the interaction of the ISM with supernovae and the winds from massive stars, rather than using cosmological simulations or those simulating the evolution of complete galaxies, which are the focus of most hydrodynamic codes.
It has also been used on larger scale outflows such as galactic filaments originating in star cluster winds \citep{Rodriguez2008} and a galactic jet from a galaxy cluster \citep{Rodriguez2006}.

This code is well suited to our problem, as it is easy to set the relevant physical parameters. We ran the code within an elongated box as this gave the best fit to the evolution of the bubble, with a maximum allowed resolution in the XYZ coordinates of 1024x1024x2048, corresponding to a physical size of 1000x1000x2000pc. The maximum resolution is just the number of cells, provided that all the cells are at maximum refinement. The time step for the outputs was set at 0.25 Myr.

The first step is to introduce the parameters which control the properties of the ISM and the injection of energy from the cluster. To model the input of momentum we assumed continuous injection and divided the total kinetic energy by the photometric age of the cluster, using a wind velocity of 1000 km s$^{-1}$. There are two implications of this choice: in the first place we are assimilating the momentum added by the supernovae to that injected by the winds. This assumption would be very inaccurate during the initial stages of the expansion, but given the current mass and radius of one of our shells it should not affect the outcome significantly. However, equating the current kinetic energy to the total input from the stars ignores radiation losses, implying that the input momentum is underestimated.

The distribution of the surrounding gas is more problematic, as we have little information on its details in order to deduce valid inputs. We have used a simple model of a disc, with an exponentially decreasing density profile in the Z direction, and isotropic XY. This gives us two free parameters, the density in the plane and the scale height. As we lack reliable measurements of these in the object under study we performed several simulations, changing the parameters in order best to match observations, until we obtained a satisfactory match with a scale height of 200 pc and a central density of 4 cm$^{-3}$. These values are not unreasonable, but they are not uniquely determined, an increase in one of them can offset a decrease in the other. The shape of the bubble would be slightly affected by these interchanges, but the effect is not easy to detect when matching the simulation to the observed bubble. The initial simulations were run assuming the thickness of the shell was 1\% of its radius, but based on the results from these we deduced that the best value to assign was about 15\% of the radius, and this value was used when correcting the quantities in Table \ref{prop_tab}. We also checked that with the revised energy and ISM distributions this thickness to radius ratio was still valid.

\subsection{Results of the modelling exercise}

\begin{figure}
\includegraphics[width=\linewidth]{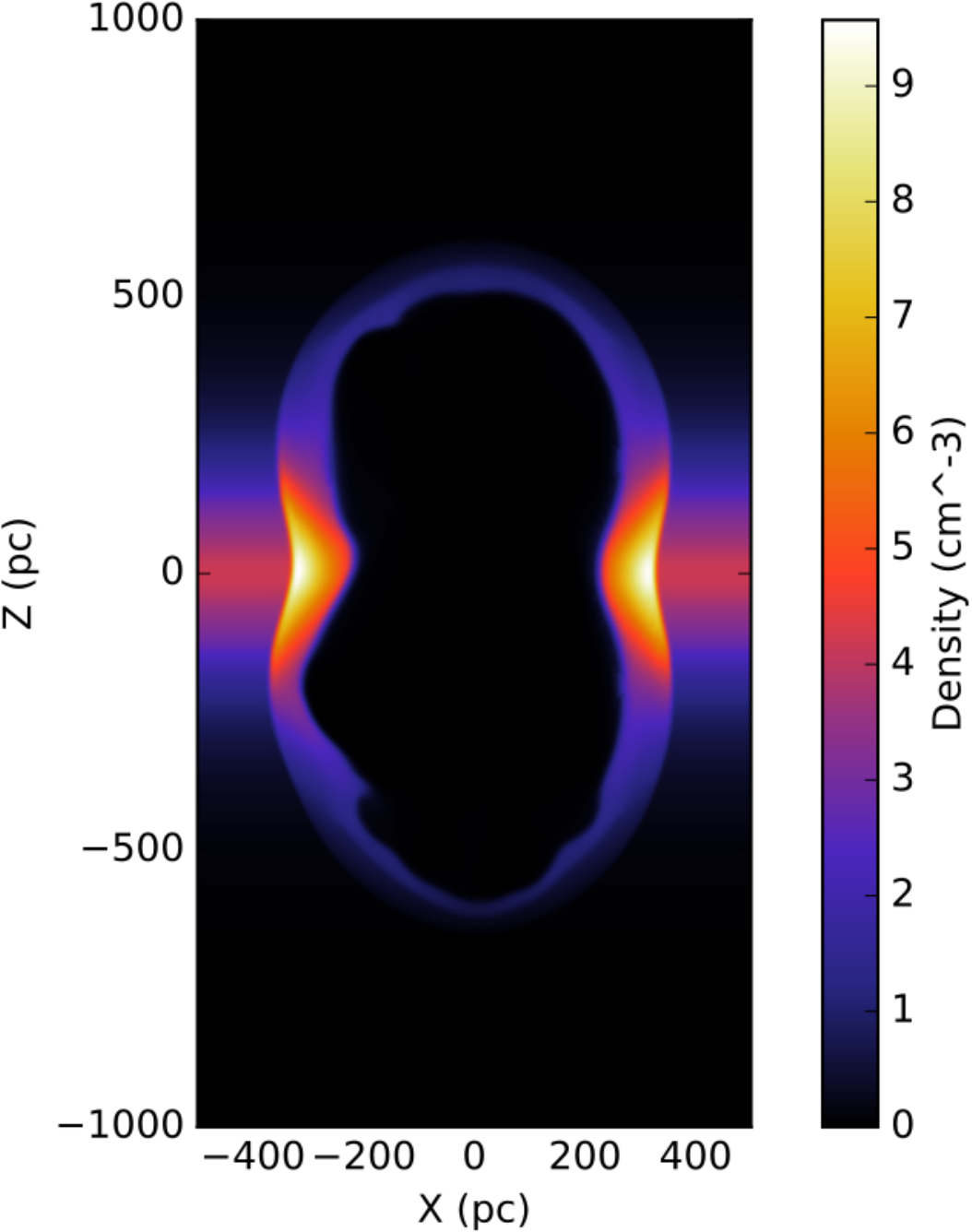}
\caption{Result of the simulation at age 4.0 Myr, the figure represents an XZ slice of the density, showing a distorted shape caused by the density gradient in the disc. The bubble is elongated in the Z axis, following the direction of decrease in density. The density in the shell is also not constant, being much higher for the part expanding into the disc.
The linear color scale gives the number density in units of cm$^{-3}$}
\label{simulation}
\end{figure}

With the model and the assumptions described in section 5.1 we can reproduce well the size and velocity of bubble 1. In Figure \ref{simulation} we show a density cross-section in the XZ plane, as well as a simulated observation in the XY plane and the observed detection for comparison. The effects of the exponential fall-off in gas density perpendicular to the plane are clear, as the bubble is elongated in the Z direction; the density of the shell is also clearly direction dependent.

\begin{figure*}
\includegraphics[width=\linewidth]{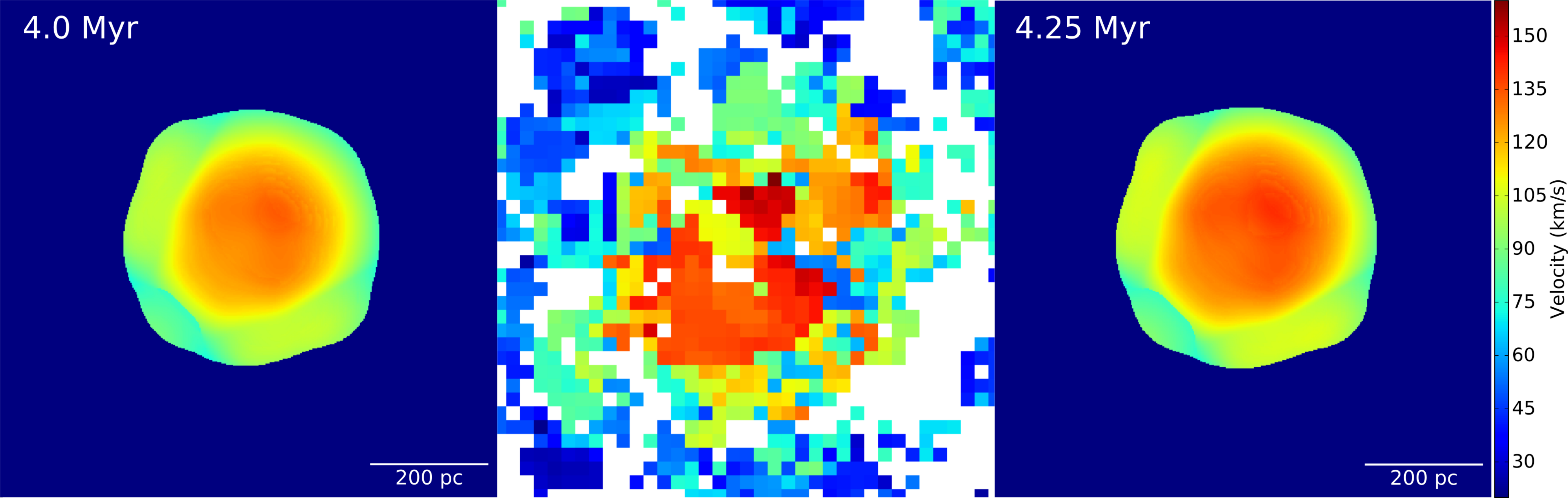}
\caption{Comparison between mock observations of the simulation bubble and the expansion map for this bubble. We show two close time-frames to show the evolution of the bubble appearance in time.}
\label{simulation2}
\end{figure*}

Figure \ref{simulation2} shows a simulated observation in the XY plane and the observed detection for comparison, obtained by projecting the velocity at the point of maximum density in the Z direction, i.e. for each X, Y position we find where the density is maximum along the Z axis and assign the corresponding velocity projected along this axis. The result is very similar to what we observe; not only is the central expansion velocity almost the same (which is the result of the inputs to this specific model) the distribution of observed velocities also follows the pattern displayed in the model of a slow decrease with projected radius before an abrupt fall to zero. This arises from the elongated shape of the superbubble, and should be characteristic of bubbles close to breakout. This model enhances the plausibility of our detection of a superbubble with the measured physical parameters, but we can further test our approximations by deriving the predicted physical parameters of the simulated superbubble, so as to compare them to the observed values.

In Table \ref{sim_tab} we show a set of parameters for the simulated bubble for two time steps, one corresponding to the closest match in expanding velocity at 4.0 Myr, close to the photometric age of the cluster (3.8 Myr) as given by \cite{Whitmore2002} which was the value we used to compute the injection rate, and another 0.25 Myr later, to compare with properties which have evolved with time.

\begin{table}
\centering
\caption{Results of the simulation}
\setlength{\tabcolsep}{5pt}
\begin{tabular}{lcccc}
\hline
	    &	& Observed	&	4.0 Myr	&	4.25 Myr	\\
Radius	& (pc) &391	&	361	&	378	\\
$v_{exp}$&(km/s)&	132	&	134	&	141	\\
$n_e$ (Z)&(cm$^{-3}$)&	0.73&	1.08&	0.97\\
$n_e$ (XY)&(cm$^{-3}$)&		&	9.49&	9.23\\
Ek	    & ($10^{51}$ erg)&	267	&	151	&	164	\\
Mass	&($10^{35}$ Kg)&	31	&	106	&	115	\\

\end{tabular}

\textbf{Notes.} The first column lists the observational properties for bubble 1, the other two correspond to two time steps in the simulation. The radius for the simulation results is taken as half the highest separation of density maxima in XY planes. The expansion velocity refers to the velocity in the Z direction at the centre of the projected bubble. $n_e$ (Z) is the maximum density along the Z axis, we consider this measurement equivalent to the measurement done on the expansion map given the area of the bubble from which we extract the properties. $n_e$ (XY) is the maximum density in the XY axes, corresponding to expansion into the disc, we do not have a measurement of this value from the observational data. It is important to note that the kinetic energy and mass cannot be compared directly as the observational quantity is overestimated for the kinetic energy and the mass is underestimated (see text).
\label{sim_tab}
\end{table}

We show also the measured parameters for ease of comparison. The density we observe coincides reasonably well with the density in the centre of the projected shell, but is quite different from the density in XY, the plane of the disc. This could be expected, as the emission we receive comes from the approaching part of the shell. However the discrepancy is carried into the computed mass of the model shell, where there is a considerable difference between this mass and the mass we measure. This is clearly important for the determination of the kinetic energy, as a mass discrepancy translates directly into the same discrepancy factor in the kinetic energy. As it happens, there is a compensating effect because the denser parts of the shell are also significantly slower, and the kinetic energy depends on the square of the velocity. We made the appropriate calculations on the simulation to ascertain the difference made because we observe only the approaching part of the shell. The result was that our observational technique underestimates the mass in the simulation by a factor 2, while the kinetic energy is overestimated by a factor 3. 
Given that we have used the observed measurement of the kinetic energy as an input to represent the total kinetic energy fed into the bubble, while according to the simulation around half the kinetic energy has been lost by radiation, the current simulation is a really reasonable fit to the data, and accomplishes the basic goal of analyzing the plausibility of the measured parameters for such a superbubble, given the limited data set for any bubble determined at the distance of the Antennae.

The relations between the input physical parameters and those found by performing a mock observation of the simulation results could be used to perfect the simulation iteratively until we obtain an exact reproduction of the observed data. While tempting, we admit that this would not enhance the scientific value of the work, as it would in practice be "overfitting" the data. The model is schematic, and among the simplifications we note that we have not taken the supernova kinetic impact directly into account but have imitated this, using winds. We have also assumed a homogeneous ionized ISM, and have not included the presence of molecular clouds or other inhomogeneities. If these could be well characterized by observation we would be able to use the observed kinematic parameters to derive accurate values for the energy injected, and thus constrain the properties of the cluster, but with the available information all we have tried to do is to present some useful semi-quantitative comparisons which put this work in a plausible context.

\section{Degree of shock-induced ionization}

There are two main mechanisms for injecting kinetic energy from stars in a massive stellar cluster into the expanding shell of a superbubble, stellar winds, and supernova explosions. Supernova shocks are far more violent than the continuous input due to fast winds, and this is reflected in the ionization state of a shell produced by each of the mechanisms. Collisional excitation is much more important in supernova remnants, while wind-blown bubbles are photoionized. By spectroscopic examination of the state of ionization of the shell we can try to determine which mechanism currently dominates its excitation, and therefore whether winds or supernovae are the principal contributors to the expansion at the stage when we observe it. For this we use the diagnostic plots from \cite{Sabbadin1977,Garcia1991} in which the line ratios [SII]/H$\alpha$ and [NII]/ H$\alpha$ are used.

The observations were made using long slit spectroscopy with the ISIS spectrograph on the WHT. The slit, with a 1 arcsec width, was placed to cover the overlap zone, including bubbles 1 and 3. A modified version of \textsc{bubbly} was run on each of the lines: [SII]$_{6716}$, [SII]$_{6734}$, [NII]$_{6584}$, and H$\alpha$. The total flux for the [NII] doublet was calculated using the relation [NII]$_{6584}$=3*[NII]$_{6548}$ which is virtually independent of the physical parameters of the gas.

\begin{figure*}
\includegraphics[width=\linewidth]{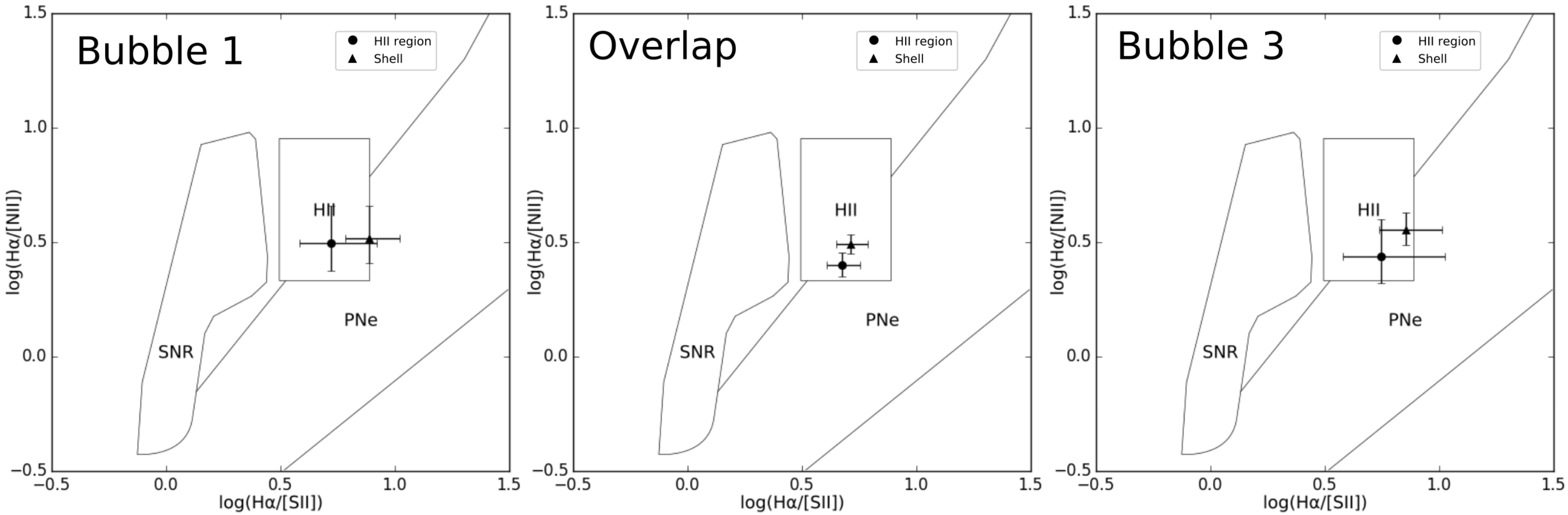}
\caption{Identification diagrams used to distinguish between supernova remnants, HII regions and planetary nebulae. We present measurements of shell (triangles) and main peak (circle) ratios for bubbles 1 and 3, as well as the overlap region. All the shells are clearly placed within the limits of the HII region area, meaning that at this stage they are primarily ionized by radiation, rather than supernova shocks. It is interesting to note, though, that they all fall closer to the supernova remnant area than their corresponding HII region emission.
}
\label{fig_ratios}
\end{figure*}

In Fig. \ref{fig_ratios} we show the results for the two bubbles and the incoherent region, in which we compare the line intensity ratios for the shells with those for the underlying HII regions. The circles correspond to the HII region flux while the triangles correspond to the shell.

The shell ratios were calculated using the fluxes in the secondary peaks, while the fluxes for the regions were obtained from the central primary peak. We can see from the figure that in all the cases both the shell of the bubble, and the HII region show ratios well within the zone of the diagram which characterizes an HII region, showing that photoionization is currently dominant compared with shock-induced ionization. The points for the shells are, however, clearly somewhat nearer to the SNR zone, which implies a somewhat greater role for shocks in the shells than in the bodies of the HII regions. This result is what one would expect, given the masses and the evolutionary stages of all our observed bubbles. A supernova remnant has masses of order a few solar masses, while our bubbles have accumulated many thousands of solar masses, as well as having had time to relax from initial supernova shocks. At this stage, the addition of an additional supernova remnant expanding into the shell would not have a strong effect on its ionization state.

\section{Summary and conclusions}

\begin{itemize}
\item We applied \textsc{bubbly}, a program sensitive to the presence of expanding shells to a Fabry-P\'{e}rot observation of the Antennae pair of galaxies mapping the H$\alpha$ emission line. The program detects and fits multiple components to the H$\alpha$ line profile, for which the data obtained with GH$\alpha$FaS is ideal given its high kinematic (8 km s$^{-1}$) and spatial (seeing-limited) resolution.

\item Using a map of the detected expansion we report the presence of 17 superbubbles associated with most of the brightest HII regions in the object. We estimate the sizes, expansion velocities and luminosities of the superbubbles and use these quantities to derive their shell density, mass, kinetic energy and age. The bubbles range between $\sim$150-500 pc in radius, with low shell densities and kinetic energy between a dozen and a few hundred supernovae. We use age determinations from the literature for the young clusters in this object and find that many of our bubbles have a matching age determination of a similar value, taking into account that the age we present is a lower limit assuming free expansion.

\item The expansion map shows not only the superbubbles but also extended detections which cover most of the galaxy and do not show enough spatial coherence to be identified as superbubbles. Nevertheless, given that their expansion velocities are generally on the low side they should not originate in noise, but rather in spatially unresolved clusters. We make an estimate of the kinetic energy contained in this unresolved expansion with several approximations and find that it is of the same order as the energy in the bubbles, meaning that the biggest HII regions do not dominate the kinetic energy budget in this object.

\item To assess the validity of our results and check for bias in the determination of the physical properties we use hydrodynamic simulations to reproduce one of our bubbles using the physical parameters as inputs. We manage to reproduce a bubble with similar observational properties. Projecting the simulation into a mock expansion map as our program would observe it shows a remarkably similar distribution of velocities due to projection effects, showing a slow decrease in observed expansion velocity before cutting off. This arises from the non-spherical shape of the bubble induced by the density profile of the disk.

\item The simulation also serves to check for bias in the physical parameter determination with our method, as we only probe the thinner, faster part of the shell which approaches us. We find that in this case the kinetic energy is overestimated by a factor 3 while the mass in underestimated by a factor 2.

\item We also use long-slit observations of HII regions along the overlap zone to probe the ionization mechanisms in the expanding shells. A diagnostic on the ratio of forbidden lines [SII]$_{6716,6734}$ and [NII]$_{6584}$ to H$\alpha$ shows that photoionization dominates this process, though the forbidden line ratios are higher for the shells rather than the bulk of the HII region, showing that collisional excitation is more important in the shells.
\end{itemize}

\section*{Acknowledgements}
This research has been supported by the Spanish Ministry of Economy and Competitiveness (MINECO) under the grant AYA2007-67625-CO2-01, and by the Instituto de Astrof\'{i}sica de Canarias under project P/308603. JEB acknowledges financial support from the DAGAL network from the People Programme (Marie Curie Actions) of the European Union's Seventh Framework Programme FP7/2007-2013/ under REA grant agreement number PITN-GA-2011-289313.
PFV and ARG thank financial support from DGAPA-PAPIIT (UNAM) grant IG100516.
MR acknowledges the grants IN 103116 from DGAPA-PAPIIT, UNAM and CY-253085 from CONACYT.

The William Herschel Telescope is operated on the island of La Palma by the Isaac Newton Group in the Spanish Observatorio del Roque de los Muchachos of the Instituto de Astrof\'{i}sica de Canarias.

     \bibliographystyle{mn2e}

\label{lastpage}

\end{document}